\documentclass{pazhb}
\usepackage[hyperfootnotes=false]{hyperref}
\usepackage{color}
\usepackage{epsf}
\usepackage[utf8]{inputenc}
\usepackage[russian]{babel}
\usepackage[T2A]{fontenc}
\usepackage{amsmath}
\usepackage{amsfonts}
\usepackage{amssymb}
\usepackage{epsf}
\usepackage{graphicx}
\usepackage{gensymb}
\usepackage{float}
\usepackage{longtable}
\usepackage{changes}
\usepackage{wrapfig}
\usepackage{subcaption}
\begin{document}

\newcommand{\obj}{Gaia~19cwm}
\newcommand{\mm}{^{\mathrm{m}}}

\journalinfo{2021}{0}{0}{1}[0]
\UDK{///}

\title{Gaia~19cwm --- an eclipsing dwarf nova of WZ~Sge type with a magnetic white dwarf}
\author{
A.I.~Kolbin\address{1,2}\email{kolbinalexander@mail.ru},
T.A.~Fatkhullin\address{1},
E.P.~Pavlenko\address{3},
M.V.~Suslikov\address{1,2},
V.Yu.~Kochkina\address{1},
N.V.~Borisov\address{1},
A.S.~Vinokurov\address{1},
A.A.~Sosnovskij\address{3},
S.S.~Panarin\address{1,2}
\addresstext{1}{Special Astrophysical Observatory of the Russian Academy of Sciences,  Nizhnii Arkhyz, Karachai-Cherkessian Republic, Russia}
\addresstext{2}{Kazan (Volga Region) Federal University, Kazan, Russia}
\addresstext{3}{Crimean Astrophysical Observatory, Nauchnyi, Crimea, Russia}
\shortauthor{Kolbin, et al.}
}


\shortauthor{Колбин А.И. и др.}

\shorttitle{Gaia~19cwm --- a dwarf nova of WZ Sge type}

\begin{abstract}
The spectral and photometric studies of the cataclysmic variable Gaia~19cwm (or ZTF19aamkwxk) have been performed. Based on the analysis of long-term variability, it is concluded that the object belongs to WZ~Sge type stars. The light curves show eclipses recurring with an orbital period of $86.32048 \pm 0.00005$~min, as well as an out-of-eclipse variability with a period of $\approx 6.45$~min.
The latter period is stable for $\sim 4$ years and appears to correspond to the rotation of a magnetic white dwarf, i.e., Gaia~19cwm is an intermediate polar. The {\obj} spectra show photospheric lines of the white dwarf, and Doppler tomograms demonstrate the presence of an accretion disk and a hot spot. Analysis of the eclipse light curve gives an estimates of the white dwarf mass $M_1 = 0.66\pm0.06$~M$_{\odot}$, the donor mass $M_2 = 0.073 \pm 0.015$ ~M$_{\odot}$, and the orbital inclination $i=83.8 \pm 1.1^{\circ}$. Modeling of the spectral energy distribution gives the white dwarf temperature of $T_{eff}\approx 13000 $~K. The X-ray luminosity $L_X = (1.6 \pm 0.3) \times 10^{31}$~erg/s allows to assign {\obj} to a small group of low-luminosity intermediate polars. 

\end{abstract}


\section{Introduction}

Cataclysmic variables are semi-detached binaries consisting of a white dwarf and a low-mass donor star \citep{Warner95}. The donor fills its Roche lobe and loses matter from the vicinity of the Lagrangian point L$_1$. When the magnetic field of a white dwarf is weak ($B<0.1$~MG), the accreted matter forms an accretion disk. The accretion stream emerging from the Lagrangian point L$_1$ forms a hot spot or, as hydrodynamic calculations show, a hot line \citep{Bisikalo03} when interacting with the accretion disk. At moderate magnetic fields ($B\sim 0.1 - 10$~MG), the internal parts of the accretion disk are destroyed and its matter flows along the magnetic field lines to the accretor. Such systems are called DQ~Her type stars or intermediate polars \citep{Patterson94}. Stronger magnetic fields ($B\sim 10 - 100$~MG) have AM~Her type systems or polars in which the accretion disk is not formed \citep{Cropper90}.

\begin{table*}[h!]
\caption{The log of {\obj} observations. The telescopes and instruments involved in the observations, dates of observations, durations of observations, number of frames obtained ($N$), spectral bands (Clear --- observations without photometric filter), spectral resolutions $\Delta \lambda$ (for spectral observations), and exposures ($\Delta t_{exp}$) are listed.}
\label{log}
\begin{center}
\begin{tabular}{lcccccc}
\hline
 Telescope/ & Date,  & Duration,        & $N$   & Band            & $\Delta \lambda$, & $\Delta t_{exp}$, \\
 Instrument                    & UT    & $\mathrm{HJD}-2460000$    &       &                   &  \AA         & s  \\\hline
BTA/SCORPIO         & 24/25 Apr. 2023   & 059.40950--059.42361   & 3     & 	3500--7200~\AA   &  12      &      600 \\
Zeiss-1000/EM       & 15/16 Apr. 2024   & 416.47349--416.50841  & 601   & Clear & --    & 5  \\
ZTSh/FLI             & 02/03 May 2024    & 433.42028--433.56291  & 516   & Clear & --    & 20 \\
ZTSh/FLI             & 03/04 May 2024    & 434.42025--434.55609  & 801   & Clear & --    & 10 \\
ZTSh/FLI             & 05/06 May 2024    & 436.40507--436.56092  & 970   & Clear & --    & 10 \\
Zeiss-1000/EM       & 16/17 May 2024    & 446.37578--446.42804  & 680   & Clear & --    & 5  \\
Zeiss-1000/EM       & 16/17 May 2024    & 446.48884--446.51660  & 240   & Clear             & --        &   10  \\
ZTSh/FLI             & 31 May/ 01 Jun. 2024& 462.31258--462.54260   & 307   & $\mathrm{I_C}$             & --        & 60 \\
ZTSh/FLI             & 01/02 Jun. 2024   & 463.31177--463.54355   & 314   & $\mathrm{I_C}$   & --        & 60 \\
ZTSh/FLI             & 02/03 Jun. 2024   & 464.30822--464.53970   & 251   & $\mathrm{I_C}$   & -- & 60 \\
ZTSh/FLI             & 03/04 Jun. 2024   & 465.30126--465.54177   & 324   & $\mathrm{I_C}$   & --        & 60 \\
BTA/SCORPIO         & 01/02 Jul. 2024   & 493.33950--493.46849  & 36    & 3600--5400~\AA   & 5.5    & 300 \\
Zeiss-1000/EM       & 03/04 Jul. 2024   & 495.28914--495.36397  & 3080  & Clear             & --        &   2    \\
  \hline
\end{tabular}
\end{center}
\end{table*}

At low accretion rates ($\dot{M}\lesssim 10^{-9}$~M$_{\odot}$/year), cataclysmic variables with a weak magnetic fields exhibit quasi-periodic outbursts with an amplitude of $2 - 7$~mag. Such systems are called dwarf novae, and their outbursts are associated with thermal instability of the accretion disk \citep{Cannizzo93}. The nature of the thermal instability lies in a sharp increase in viscosity (or $\alpha$ parameter in the model of \citealt{Shakura73}) that occurs during the ionization of hydrogen after a certain amount of matter has been accumulated by the disk. It is assumed that the disk's viscosity parameter is $\alpha \sim 0.01$ in a quiescent state and $4-10$ times higher during an outburst. Among dwarf novae, stars of the SU~UMa type can be distinguished. These stars exhibit outbursts with an amplitude of $2-5$~mag and a duration of several days, which are called normal outbursts. In addition to normal outbursts, so-called superoutbursts are also observed, which are larger in amplitude (by $1-2$~mag) and duration (about two weeks). The occurrence of superoutbursts is explained by the model of tidal-thermal instability \citep{Osaki95}, which takes into account, in addition to the thermal instability mentioned above, the tidal instability that occurs when the accretion disk reaches a resonance radius of $3:1$. A feature of superoutbursts are the so-called superhumps --- small variations in brightness with a period $P_{sh}$, several percent larger than the orbital period $P_{orb}$. It is assumed that during a superoutburst the accretion disk is elongated and slowly precesses in the direction of the donor's orbital motion. As a result of the tidal interaction of the edges of the elongated disk with the donor, brightenings (i.e., superhumps) occur that repeat with the beat period of the precessional and orbital motion.

Among SU~UMa type stars, a subclass of WZ~Sge type stars is distinguished, which exhibit only superoutbursts with an amplitude of $6-8$~mag, repeating at intervals (the so-called supercycle) of $\sim 10$~years (see more about WZ~Sge type stars \citealt{Kato15}). The absence or rarity of normal outbursts, as well as the long duration of the supercycle in WZ~Sge type stars, currently has no unambiguous interpretation. To reproduce the outburst activity, the tidal-thermal instability theory requires a very low viscosity of the disk ($\alpha \lesssim 10^{-4}$) that is difficult to explain. The activity of WZ~Sge stars can also be explained within the framework of ``standard'' viscosity values, but this requires a limitation of the inner radius of the accretion disk due to the evaporation or magnetic field of the white dwarf \citep{Hameury97, Matthews07}.

The variable Gaia~19cwm (ZTF19aamkwxk, AT2019kwk; $\alpha_{2000} = 16^h 27^m16.77^s$, $\delta_{2000} = +04^{\circ}06' 02.58''$) was discovered as an optical transient by the Gaia space observatory. Based on the analysis of ZTF data, \cite{Szkody21a} included {\obj} in candidates for cataclysmic variables. From the analysis of the ZTF data, we concluded that {\obj} could be a WZ~Sge type star. We also discovered eclipses, which makes {\obj} a very interesting object for studying the nature of WZ~Sge stars (in the list \cite{Kato15} there are only four eclipsing systems out of almost a hundred WZ~Sge stars). In addition, the ZTF data showed signs of out-of-eclipse variability, which could be related to the rotation of the white dwarf or with its non-radial pulsations. All this makes {\obj} an interesting object for detailed photometric and spectral studies, the results of which are presented in this paper.

\section{2. Observations and data reduction}

\subsection{Photometry}

The photometric observations of {\obj} were carried out on the 1-m Zeiss-1000 telescope of the Special Astrophysical Observatory of the Russian Academy of Sciences (SAO RAS). The telescope was equipped with the Andor iXon Ultra 888 EMCCD detector. Also, observations of {\obj} were carried out on the 2.6-m ZTSh telescope of the Crimean Astrophysical Observatory of the Russian Academy of Sciences, equipped with the FLI PL-4240 photometer with an E2V 42-40 CCD\footnote{See the website http://crao.ru/ru/telescopes for more details}. The log of photometric observations is included to the Table \ref{log}.

The observational data were processed using conventional technique, including bias subtraction, flat-fielding, and cosmic ray removal. Aperture photometry was performed using the photutils library\footnote{The photutils library is available at https://photutils.readthedocs.io/en/stable/.}. 


\subsection{Spectroscopy}

A set of {\obj} spectra was obtained on the nights of April 24/25, 2023 and July 01/02, 2024 at the 6-m BTA telescope of SAO RAS. The telescope was equipped with the SCORPIO\footnote{A description of the SCORPIO focal reducer can be found at https://www.sao.ru/hq/lsfvo/devices/scorpio/scorpio.html} focal reducer, which was used in long-slit spectroscopy mode \citep{Afan05}. On the night of April 24/25, 2023, a volumetric phase holographic grating VPHG550G was used as a disperser, which allowed to cover the spectral range $3900-7500$~\AA. The observations were made with a slit width of $1.2''$, giving a spectral resolution of $\Delta \lambda \approx 12$~\AA. On the night of July 01/02, 2024, observations were carried out using the VPHG1200B grism with a working range of 3600--5400 ~\AA\, and a resolution of 5.5~\AA\, (with a slit width of $1.2''$). The standard stars HZ44 and AGK+81$^{\circ}$266 were observed to perform the spectrophotometric calibration. The processing of the obtained data was carried out in the IRAF environment\footnote{The IRAF astronomical data processing and analysis package is available at https://iraf-community.github.io.}, following the standard methodology of the long-slit spectroscopy data reduction.

\section{3. Analysis of Photometry}

\begin{figure*}[h]
  \centering
	\includegraphics[width=\textwidth]{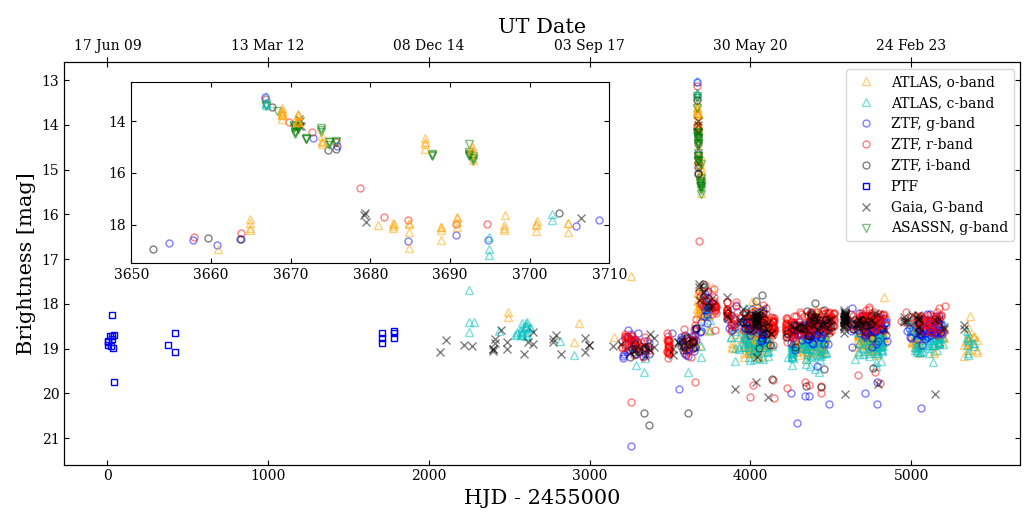}
\caption{Long-term light curve of {\obj}, reconstructed from ATLAS ($o$, $c$), ZTF ($g$, $r$, $i$), PTF, Gaia ($G$), ASAS-SN ($g$) and Pan-STARRS ($g$, $r$, $i$, $z$, $y$) data.}
\label{fig:long_term}
\end{figure*}

To analyze the long-term variability of {\obj}, archival data from the  ZTF (Zwicky Transient Facility, \citealt{masci18}), Gaia space observatory, ATLAS\footnote{ATLAS photometry is available at https://fallingstar-data.com/forcedphot. ATLAS observations are performed in the $o$ (560–820~nm) and $c$ (420–650~nm) passbands.} (Asteroid Terrestrial-impact Last Alert System, \citealt{Tonry18}), Pan-STARRS (Panoramic Survey Telescope and Rapid Response System, \citealt{ps}), PTF (Palomar Transient Factory, \citealt{Low09}), and ASAS-SN (All Sky Automated Survey for SuperNovae, \citealt{Shappee14}) surveys were used. The resulting light curve, covering $\approx 15$~years of observations, is shown in Fig. \ref{fig:long_term}. It shows an outburst (probably a superuotburst) with an amplitude of $\approx 6$~mag and a duration of about $12-14.5$~days. The onset of the outburst lies between epochs $\mathrm{HJD} = 2458663.79$ and $\mathrm{HJD} = 2458666.64$ (i.e. between June 29 and July 2, 2019). After the outburst, two brightenings were detected around $\mathrm{HJD}=2458686$ and $\mathrm{HJD} = 2458692$, i.e. $\approx 7$ and $\approx 13$ days after the end of the outburst, respectively.

The orbital period $P_{orb} = 86.32048 \pm 0.00005$~min was determined using ZTF $g$, $r$ data after the outburst (the error was estimated using the Monte Carlo method). The period was calculated using the Lomb–Scargle method (6 harmonics were used to fit the eclipse). The ephemeris for the mid-eclipse was obtained:
\begin{equation}
    \mathrm{HJD_{min}} = 2460000.0568(1) + 0.05994478(4) \times E.
\label{ephem}
\end{equation}

\begin{figure*}[h]
  \centering
	\includegraphics[width=\textwidth]{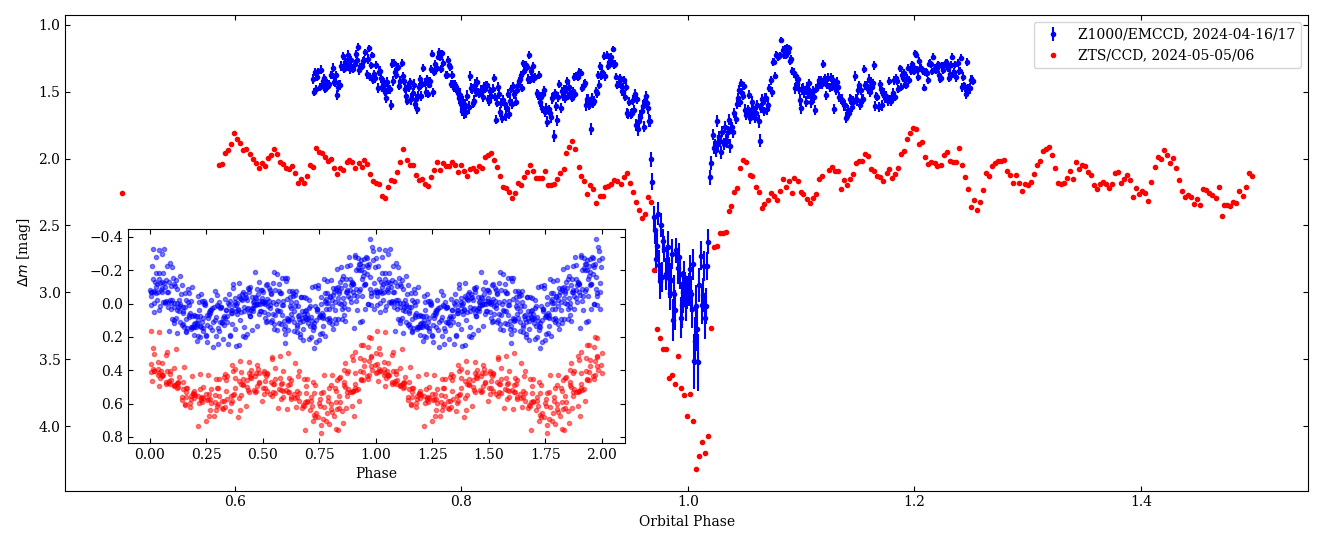}
\caption{The examples of {\obj} light curves  obtained with the Zeiss-1000 and ZTSh telescopes. Out-of-eclipse light curves folded with a period of $P_s=6.45$~min are also shown in the inset plot. The ZTSh light curve is shifted down by $\approx 0.5$~mag.}
\label{fig:lcs}
\end{figure*}

\begin{figure}[h!]
  \centering
	\includegraphics[width=\linewidth]{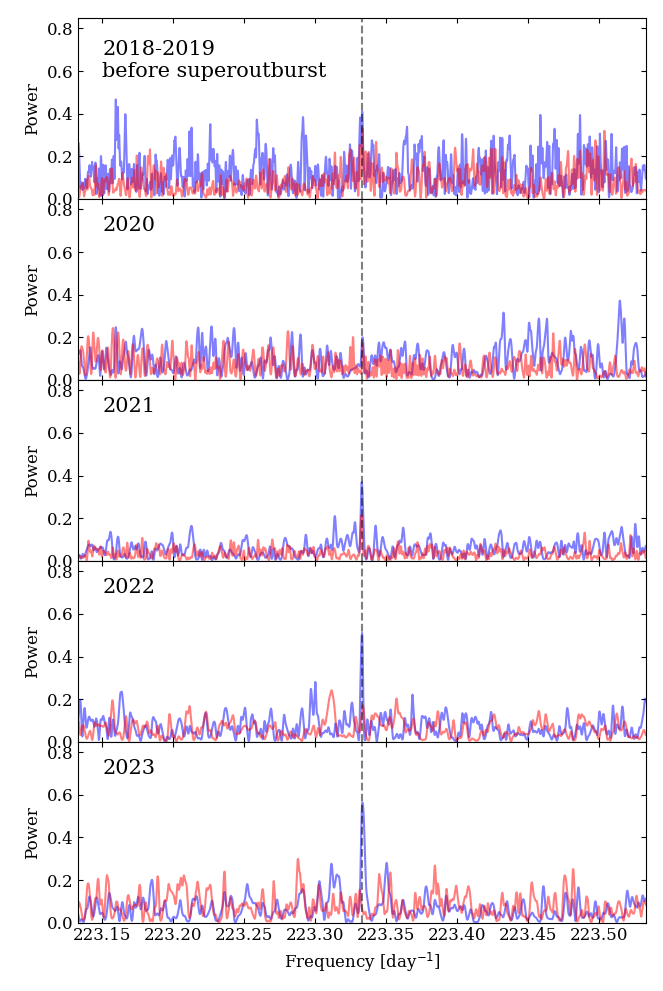}
\caption{Lomb-Scargle periodograms obtained from the ZTF data for 2018-2019 before the superoutburst, 2020, 2021, 2022, 2023. The blue line shows the periodograms constructed from $g$ data, and the red lines show periodograms for  $r$ data. The dotted line indicates the frequency $f_s = 223.323 \pm 0.001$~day$^{-1}$, found from the ZTSh observations in 2024.}
\label{fig:ztf_periods}
\end{figure}

Observations of the Zeiss-1000 and ZTSh telescopes show an out-of-eclipse variability with a period of $P_s\approx 6.45$~min. Examples of light curves showing this variability are presented in Fig. \ref{fig:lcs}. The light curves folded with the found period exhibit a double-peaked structure with a peak separation of $\approx P_s/2$. Analysis of ZTF data obtained after the outburst revealed the same periodicity. Using the Lomb–Scargle method, the period $P_s=6.4477728 \pm 0.0000006$~min (frequency $f = 223.33293 \pm 0.00002$~day$^{-1}$) was determined. Periodograms of {\obj} constructed from ZTF observations in 2018-2019 (before the outburst), 2020, 2021, 2022, and 2023 are shown in Fig. \ref{fig:ztf_periods}. In the periodograms for the years 2021, 2022, and 2023, a significant power peak is observed at a period that is consistent with $P_s$ within the errors. Thus, the period is stable for at least four years, indicating that it is related to the rotation of the white dwarf rather than to non-radial pulsations. In the latter case, one would expect the period to change as the white dwarf cools (see, for example, \citealt{Szkody21}).

\begin{figure*}
\centering
\begin{subfigure}{.49\textwidth}
    \centering
    \includegraphics[width=.95\linewidth]{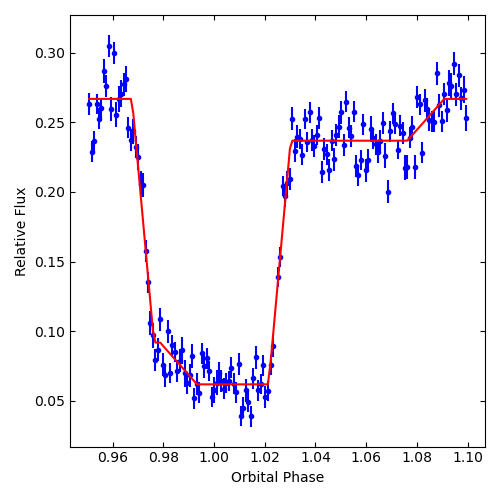}
    \caption{}
\end{subfigure}
\begin{subfigure}{.49\textwidth}
    \centering
    \includegraphics[width=.95\linewidth]{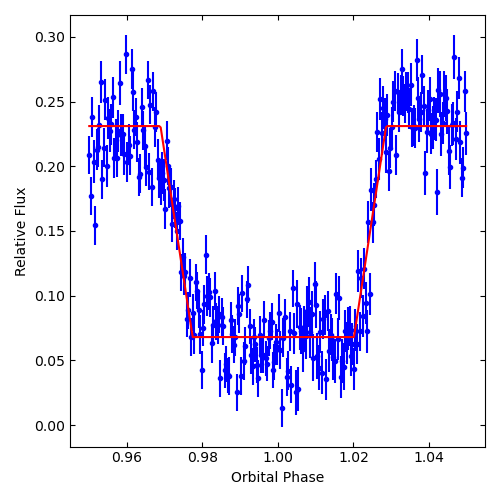}
    \caption{}
\end{subfigure}
\caption{Light curves of the eclipse of {\obj} obtained at Zeiss-1000 telescope with EMCCD on April 15/16, 2024 (a) and July 3/4, 2024 (b). The red lines show the approximating trapezoids.}
\label{figure:eclipses}
\end{figure*}

Fig. \ref{figure:eclipses}a and \ref{figure:eclipses}b show the light curves of {\obj} constructed from observations with the Zeiss-1000/EMCCD. The rotational variability was removed from the light curves by approximating the out-of-eclipse parts with a trigonometric polynomial. The first light curve exhibits a complex structure, where, in addition to the white dwarf eclipse, a weak eclipse of the hot spot is also visible. The light curve is well approximated by a set of two trapezoids. The hot spot eclipse is centered at phase $\varphi_b = 1.0359 \pm 0.0015$ and has a width (FWHM) of $\Delta \varphi_b = 0.098 \pm 0.008$. The width (FWHM) of the white dwarf eclipse $\Delta \varphi_w = 0.0510 \pm 0.0007$ and the duration of the ingress (and egress) $\Delta \varphi_{egr} = 0.0079  \pm  0.0007$ were determined from the second light curve, obtained with higher time resolution. The hot spot eclipse is no longer visible in this curve.

\begin{figure}
  \centering
	\includegraphics[width=\linewidth]{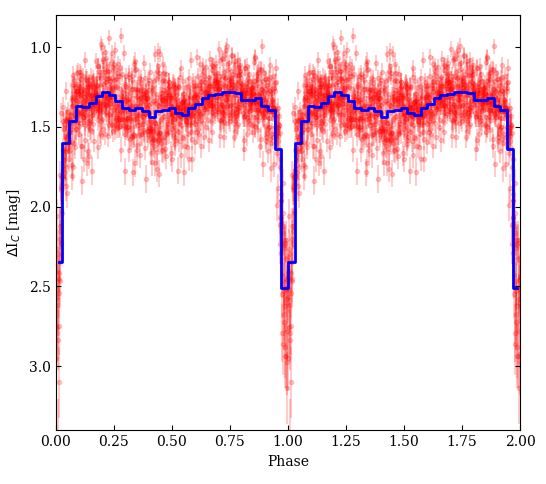}
\caption{The light curve of {\obj} folded with the orbital period (red dots). The observations were carried out in the $\mathrm{I_C}$ filter at the ZTSh telescope over four nights. The blue line shows the average of the light curve within 35 phase segments.}
\label{fig:lc_zts}
\end{figure}

In addition to the noted rotational variability, the light curves of {\obj} also show out-of-eclipse variability modulated with the orbital period. Over the orbital cycle, it has a double-peaked structure, which is evident in ZTSh observations (see Fig. \ref{fig:lc_zts}). The amplitude of the out-of-eclipse variability is $\Delta \mathrm{I_C} \approx 0.15$~mag. Possible interpretations of this phenomenon are discussed in the ``Discussion and Conclusion'' section.

\section{4. Spectral analysis}

\begin{figure*}
  \centering
	\includegraphics[width=\textwidth]{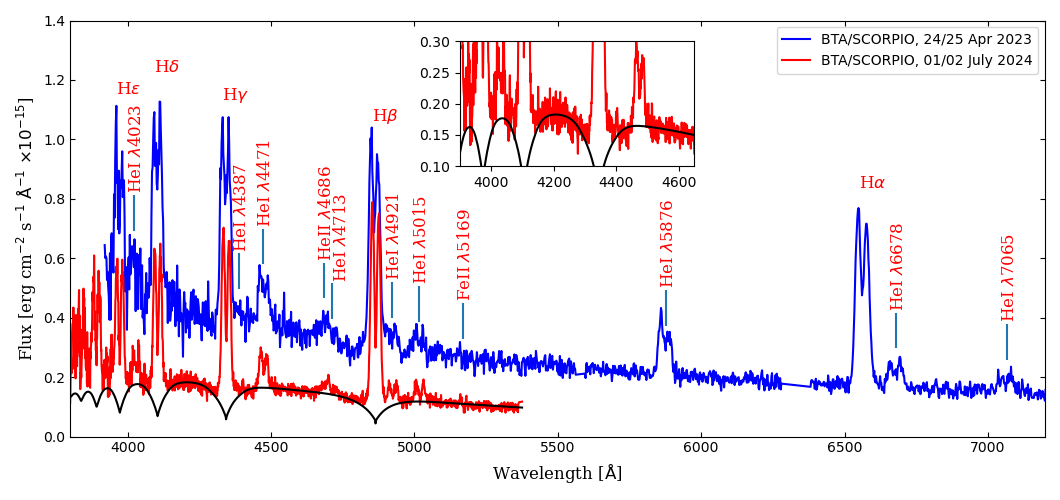}
\caption{Average spectra of {\obj} obtained from observations on April 24/25, 2023 and July 01/02, 2024 (the 2023 spectrum is shifted relative to the 2024 spectrum). The black line shows the theoretical spectrum of a white dwarf with the effective temperature $T_{eff} = 13000$~K and the surface gravity $\log g = 8.0$. The inset figure shows the broad Balmer absorptions near the region of the H$\gamma$ and H$\delta$ lines.}
\label{fig:spectrum}
\end{figure*}

The averaged spectra of {\obj} are shown in Fig. \ref{fig:spectrum}. They contain a typical set of spectral lines for cataclysmic variables: hydrogen Balmer lines (H$\alpha$, H$\beta$, H$\gamma$, H$\delta$, H$\epsilon$, H$\zeta$), neutral helium lines (HeI~$\lambda4020$, $\lambda4387$, $\lambda4471$, $\lambda4921$, $\lambda5015$, $\lambda5876$, $\lambda6678$, $\lambda7065$), a weak ionized helium line HeII~$\lambda4686$, and an ionized iron line FeII~$\lambda5169$. The hydrogen and neutral helium lines have a double-peaked structure, typical for dwarf novae with high orbital inclination ($i\sim90^{\circ}$). Around the H$\gamma$ and H$\delta$ emission lines in the spectrum from July 1/2, 2024, signs of broad absorptions formed in the white dwarf's atmosphere are observed. For clarity, this spectrum is compared with a theoretical spectrum of a white dwarf with a temperature $T_{eff}=13000$~K and a surface gravity $\log g = 8.0$ \citep{Koester10}.

\begin{figure*}
    \includegraphics[width=\columnwidth]{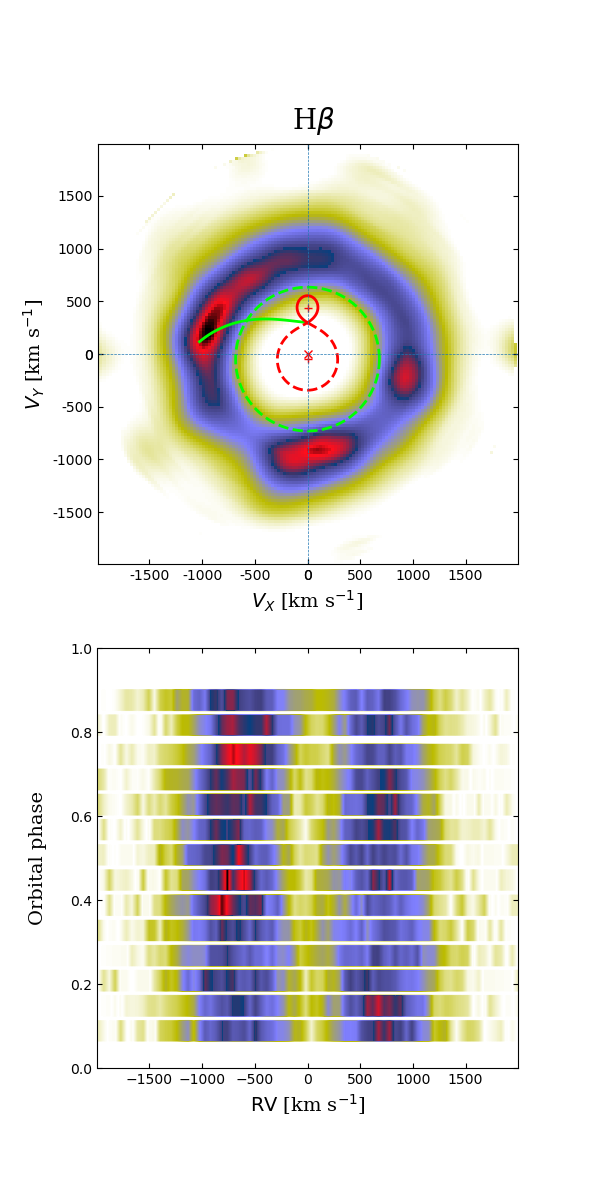}
    \includegraphics[width=\columnwidth]{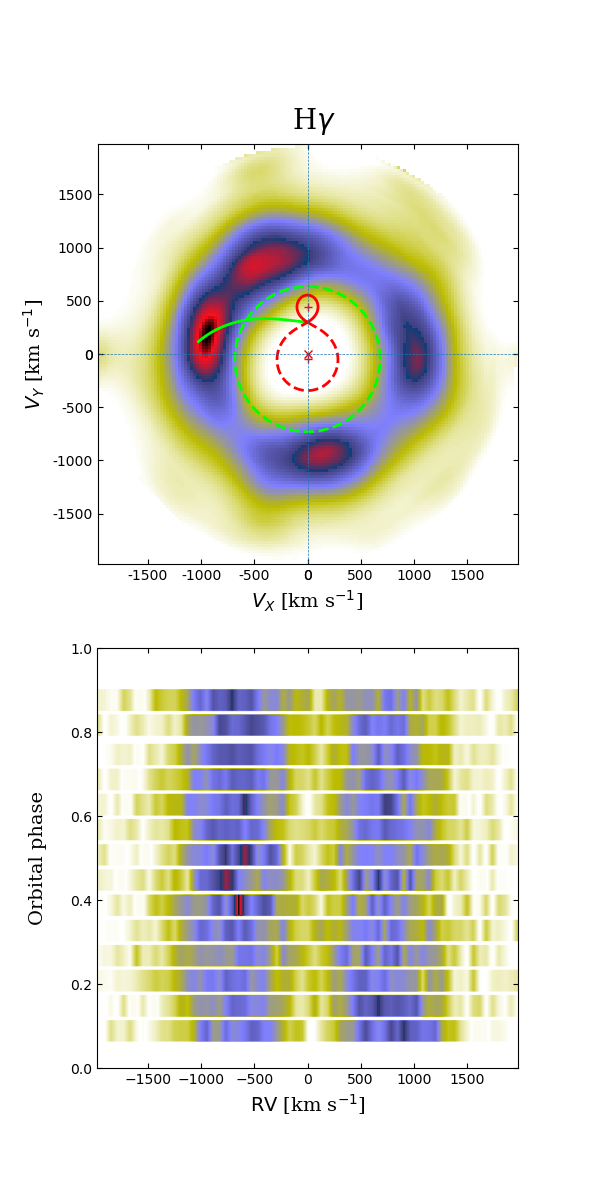}
\caption{Top panel: Doppler tomograms of {\obj} in the H$\beta$ (left) and H$\gamma$ (right) lines. The velocities of the Roche lobes of the primary (closed red dotted line) and secondary (closed solid line) are shown. The green solid line indicates the ballistic trajectory of particles escaping from the Lagrangian point L$_1$. The green dotted circle is the velocities of particles at the orbit around the white dwarf where the $3:1$ resonance is realized. Bottom panel: dynamic spectra of the H$\beta$ (left) and H$\gamma$ (right) lines used to reconstruct the Doppler tomograms.}
\label{fig:tomograms}
\end{figure*}

The Doppler tomograms of {\obj} in the H$\beta$ and H$\gamma$ lines were reconstructed from observations on July 1/2, 2024, and are presented in Fig. \ref{fig:tomograms} (for the details on Doppler tomography and the interpretation of the Doppler tomograms we refer to \citealt{Marsh16}). The tomograms were reconstructed using the doptomog code \citep{Kotze15, Kotze16}, which implements the maximum entropy method. The tomograms show a ring-shape structure corresponding to the emission from the accretion disk. A hot spot is also visible in the second quadrant of the tomograms ($v_x\le0$, $v_y\ge 0$). Both tomograms also contain a second, weaker spot in the fourth quadrant ($v_x\ge0$, $v_y \le 0$). This structure is not a common feature of dwarf novae but has been observed in some WZ~Sge type stars \citep{Alives10, Zharikov13}. A model of {\obj} with a mass ratio $q=0.11$, a white dwarf mass $M_1 = 0.66M_{\odot}$, and an orbital inclination $i=84^{\circ}$ (see the next section) is superimposed on the tomograms. It can be seen that the accretion disk in the hydrogen lines does not exceed the $3:1$ resonance radius, and the ballistic trajectory of particles escaping from the Lagrangian point L$_1$ intersects the hot spot.

\begin{figure}
    \includegraphics[width=\columnwidth]{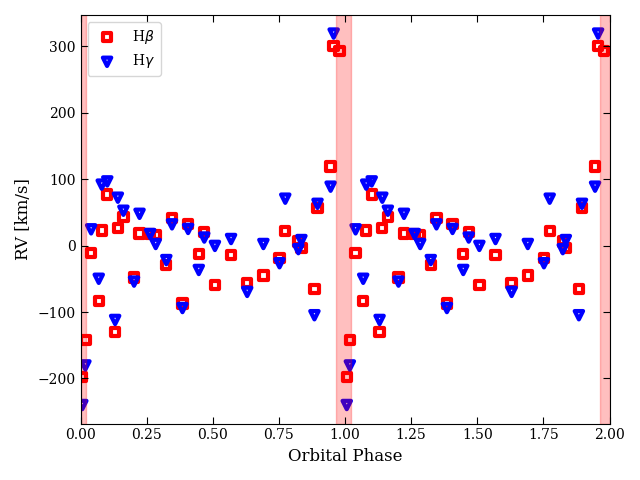}
  \caption{Radial velocity curve of {\obj} in the H$\beta$ and H$\gamma$ lines. The vertical stripes indicate the eclipse phases of the white dwarf.}
\label{fig:rvs}
\end{figure}

In the study of cataclysmic variables, the method of determining radial velocities proposed by \cite{shafter83} is often used. In this method, radial velocities are measured from the wings of emission lines, which are formed in the inner parts of the accretion disk. It can be expected that the radial velocities obtained in this way reflect the motion of the white dwarf. The determination of radial velocities $v_r$ is done by solving the equation$\int k(v - v_r) r(v) dv = 0$, where $r(v)$ is the observed profile of the spectral line in the velocity scale, and $k(v)$ is the difference of two Gaussians centered on the opposite wings of the lines. Thus, the radial velocity is adjusted so that the Gaussians contain the same amount of flux in the wings of the lines. The separation of the Gaussians was chosen  to be greater than $2 K_{spot}$, where $K_{spot} \approx 900$~km/s is the semi-amplitude of the radial velocity of the hot spot (see the Doppler tomograms in Fig. \ref{fig:tomograms}), and was limited from above by the $\mathrm{FWZI}$ of the line. The widths of the Gaussians varied from $\mathrm{FWHM}/4$ to $\mathrm{FWHM}/2$, where $\mathrm{FWHM}\approx700$~km/s is the width of the spectral line peak. Unfortunately, varying the separation and widths of the Gaussians did not yield the expected radial velocity curve for the white dwarf. Apparently, the available signal-to-noise ratio is insufficient to recover the radial velocities of the parts of the accretion disk that reflect the motion of the white dwarf. Alternatively, such parts may be entirely absent due to the limitation of the inner radius of the accretion disk by the white dwarf's magnetic field. Fig. \ref{fig:rvs} shows the radial velocity curves of {\obj} in the H$\beta$ and H$\gamma$ lines, obtained with parameters $a=2500$~km/s and $\sigma = 300$~km/s ($\sigma$  is the standard deviation of the Gaussian). The effect of the accretion disk eclipse is well pronounced, manifesting as a sharp jump in radial velocity before the center of the eclipse ($\varphi=1$) and a similar negative jump after the center of the eclipse. Outside the eclipse, the radial velocities are randomly distributed. The average error in measuring the radial velocity outside the eclipse is $\approx 25$~km/s.

\section{5. Parameters estimation}

The mass ratio $q=M_2/M_1$, the orbital inclination $i$, and the accretion disk radius $R_D/A$ (i.e. in units of semi-major axis $A$) can be estimated from the eclipse light curves of the white dwarf and the hot spot \citep{Wood89}. The width of the white dwarf eclipse $\Delta \varphi_w$ depends on the orbital inclination $i$ and the mass ratio $q$. If we assume that the hot spot is formed in the region of the intersection of the ballistic trajectory of the particles emitted from the Lagrange point L$_1$ with the outer edge of the accretion disk, then from the position of the center $\varphi_b$ and the width $\Delta \varphi_b$ of the hot spot eclipse we obtain two more equations for the parameters $q$, $i$, $R_D/A$. To calculate the $\Delta \varphi_w$, $\varphi_b$ and $\Delta \varphi_b$ grids, we used a simple cataclysmic variable model with a donor filling its Roche lobe. The hot spot was assumed to be point-like and located at the intersection of the accretion stream with the outer edge of the accretion disk. The trajectory of the accretion stream was calculated by solving the restricted three-body problem for particles escaping from the Lagrange point L$_1$ with a low initial velocity \citep{Flannery75}.

\begin{figure}
    \centering
    \includegraphics[width=\columnwidth]{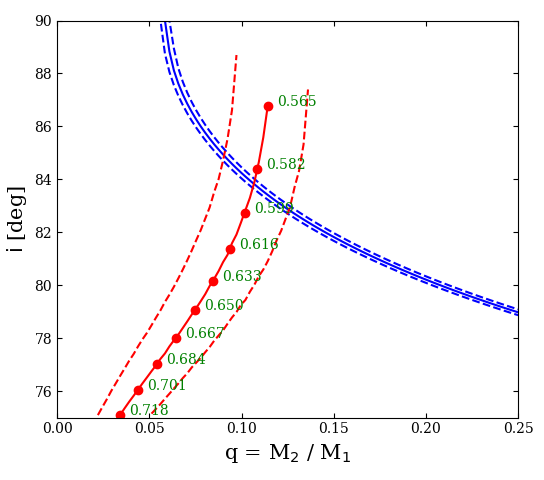}
\caption{The $i-q$ plane with solutions providing the observed width of the white dwarf eclipse (blue line) and the parameters (position and width) of the hot spot eclipse (red line). The green inscriptions indicate the disk radii in units of the major semi-axis $R_D/A$ corresponding to the position and width of the hot spot eclipse.}
\label{fig:solution}
\end{figure}

Fig. \ref{fig:solution} shows the solution obtained from the eclipse width of the white dwarf $\Delta \varphi_w$. It can be seen that it imposes a constraint on the mass ratio $q \ge 0.06$. The same figure shows the solution providing the observed center $\varphi_b$ and width $\Delta \varphi_b$ of the hot spot eclipse. It can be seen that the two solutions intersect at the mass ratio $q = 0.11 \pm 0.02$ and the orbital inclination $i = 83.8 \pm 1.1^{\circ}$. The radius of the accretion disk $R_D/A \approx 0.59$ corresponds to this solution. The egress (or ingress) duration $\Delta \varphi_{egr} = 0.0079 \pm 0.0007$ (see Section 3) corresponds to the radius of the white dwarf $R_1 = 0.0109-0.0124$~R$_{\odot}$ or, according to the relation of \cite{Nauenberg72}, to the mass $M_1 = 0.66 \pm 0.06$~M$_{\odot}$. Note that the brightness distribution over the disk of the white dwarf should be complicated due to the presence of accretion spots. For this reason, we did not use the binary model to simulate the eclipse profile, and restricted ourselves to a simpler technique assuming a trapezoidal eclipse shape. 

\begin{figure*}
\begin{subfigure}{.49\textwidth}
    \centering
    \includegraphics[width=\columnwidth]{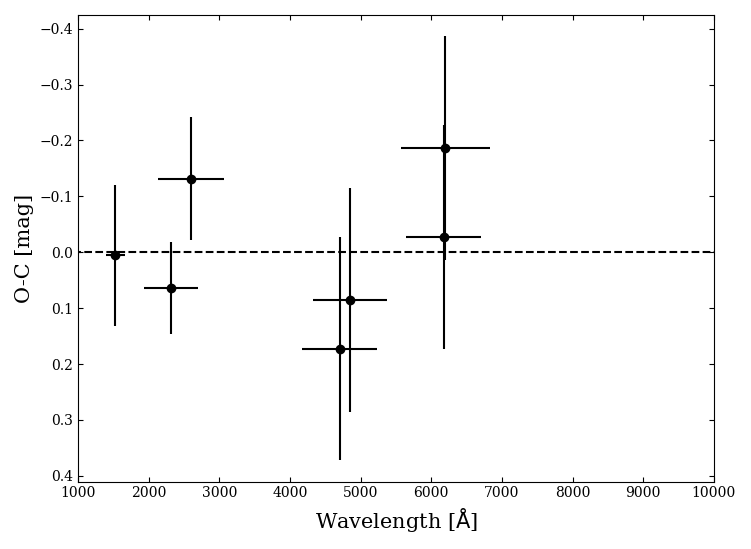}
    \caption{}
\end{subfigure}
\begin{subfigure}{.49\textwidth}
    \includegraphics[width=0.9\columnwidth]{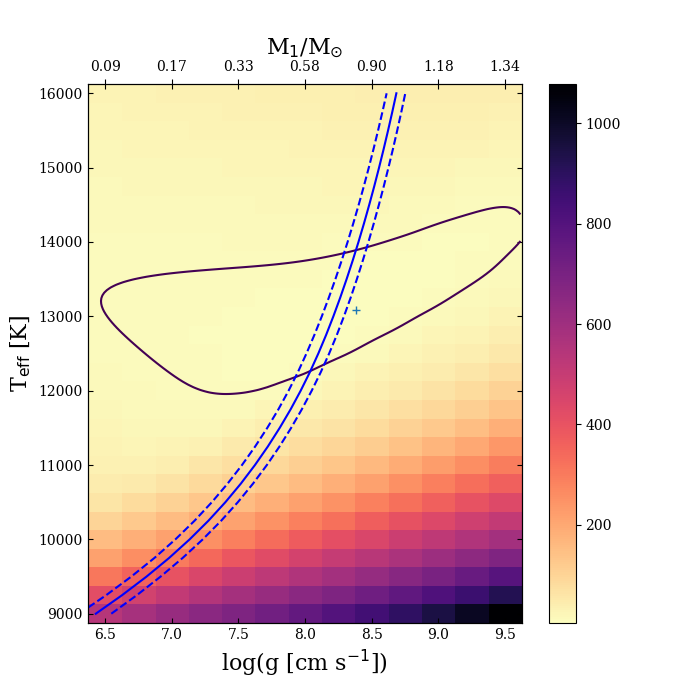}
    \caption{}
\end{subfigure}
\caption{a) The $O-C$ diagram for the observed and theoretical magnitudes of the white dwarf (horizontal bars indicate the widths of the photometric bands). b) The map of $\chi^2$ distribution in $T_{eff}-\log g$ plane. }
\label{fig:sed}
\end{figure*}

To estimate the parameters of the white dwarf's atmosphere, we used ultraviolet and optical fluxes of {\obj}. We borrowed archival observations from the GALEX observatory \citep{Boselli11} in the FUV ($\lambda_{\mathrm{eff}} \approx 1550$~\AA) and NUV ($\lambda_{\mathrm{eff}} \approx 2300$~\AA) bands, data from the Pan-STARRS \citep{Chambers16} survey, and the minimum out-of-eclipse fluxes from the ZTF survey in the $g$, $r$ bands obtained before the outburst. We also measured the fluxes in the UVW1 band ($\lambda_{\mathrm{eff}} \approx 2680$~\AA) based on observations from the UVOT telescope aboard the Swift orbital observatory ($\mathrm{id}=00014862002$). Measurements capturing the eclipse of the white dwarf were excluded from the Swift/UVOT data. It was assumed that the fluxes in the bands listed are dominated by the white dwarf radiation. Thus, the light curve of the eclipse (see Fig. \ref{figure:eclipses}a) shows that the contribution of the hot spot to the integral emission of the system is $\sim 10\%$. The eclipse of the accretion disk does not appear in the available light curves and its contribution should be smaller. The far-ultraviolet band is often used to analyze the emission of a white dwarf, where the contribution of other components of the system can be neglected. The measurements in the NUV and UVW1 bands do not deviate significantly in the $O-C$ diagram (see Fig. \ref{fig:sed}a) and it is likely that the contribution of the accretion disk and hot spot in these bands is not greater than or comparable to the corresponding contribution in the optical range. We did not use photometric data in the red part of the optical spectrum and before the Balmer jump because of the possible large contribution of the accretion disk and hot spot (see, e.g., \citealt{Pala19, Neustroev23}). According to the three-dimensional STILISM\footnote{https://stilism.obspm.fr, see \cite{extmap1, extmap3, extmap2} for details.} interstellar absorption maps, the color excess for {\obj} $E(B-V) = 0.051^{+0.037}_{-0.039}$~mag with the corresponding absorption $A_V \approx 3.1 E(B-V) = 0.16 \pm 0.11$~mag. The observed fluxes {\obj} have been corrected for interstellar extinction according to the absorption curve of \cite{Fitzpatrick99}.

The fluxes in the optical and ultraviolet bands were fitted by LTE spectra of hydrogen (DA) atmospheres of white dwarfs calculated by \cite{Koester10}. The fluxes in the photometric bands were calculated by integrating the theoretical spectra within the transmission functions of the filters\footnote{The transmission functions of the filters used are available at http://svo2.cab.inta-csic.es/theory/fps/.}. The shape of the theoretical spectrum depends on the effective temperature $T_{eff}$ and the surface gravity $\log g$. By fitting the observations, a scaling factor to the theoretical spectra that minimizes $\chi^2$ was found for each pair ($T_{eff}$, $\log g$). Since the theoretical fluxes are calculated near the surface of the star, the multiplier for the observed fluxes corrected for interstellar absorption is $\theta^2/4$, where $\theta$ is the angular diameter of the white dwarf. The map of the $\chi^2$ distribution in the $T_{eff} - \log g$ plane is shown in Fig. \ref{fig:sed}b. An isoline for a 90\% confidence level is drawn, showing the high uncertainty in $\log g$. The minimum of $\chi^2$ is reached near $T_{eff} = 13080$~K and $\log g = 8.4$ at $\chi^2_{\nu} = 0.93$ ($\nu=3$). 

The parallax $p'' = (4.21 \pm 0.24)\times 10^{-3}$ for {\obj} from the Gaia DR3 catalog \citep{Gaia20} gives the distance $D= 237\pm13$~pc. This allows us to impose additional constraints on the $\log g$ of the white dwarf (see \citealt{Kolbin24a} for details on the method used). Indeed, when fitting photometric fluxes in the ultraviolet and optical ranges, the angular diameters of $\theta$ were found. On the other hand, the angular diameter at a known distance depends only on the radius of the white dwarf, which in turn is related to $\log g$ by the relation of \cite{Nauenberg72}. The requirement that the photometric angular diameters are equil to $2 R_1(\log g)/D$ gives the solution in the $T_{eff} - \log g$ plane shown in Fig. \ref{fig:sed}. This solution gives a constraint on the surface gravity $\log g \in 8.0-8.4$ corresponding to a mass $M_1 \in 0.58 - 0.84$~M$_{\odot}$. The derived mass constraint agrees well with the estimate of the white dwarf mass $M_1 = 0.66 \pm 0.06$~M$_{\odot}$ found from the eclipse analysis. 

\section{6. The X-ray spectrum}

The variable {\obj} fell within the field of view of the XRT X-ray telescope aboard the Swift orbital observatory. The spectrum of {\obj} was obtained in the energy range $E = 0.2-10$~keV with an exposure of 3.6~ks (ObsID~00014862002). The extraction of spectral observations was performed using the online Swift/XRT data extraction service\footnote{The Swift/XRT data service is available at https://www.swift.ac.uk/user\_objects/.}  \citep{Evans09}. The spectral data were analyzed using the Xspec program of the HEASoft\footnote{The HEASoft software package is available at https://heasarc.gsfc.nasa.gov/lheasoft/.} package. The spectrum was grouped such that each spectral channel contained at least 1 count. To estimate the X-ray luminosity, the spectrum was modeled using an absorbed bremsstrahlung model (tbabs(bremss) in Xspec). The tbabs absorption model used the chemical abundances presented in \cite{Wilms00}. Applying the empirical relation from  \cite{Foight16} to the interstellar absorption estimate obtained in the previous section, we found a hydrogen column density of $N_H = (0.45 \pm 0.018) \times 10^{21}$ cm$^{-2}$, which was fixed during the modeling of the X-ray observations. Since the spectrum had small number of counts (only 18 counts), the fitting was performed using the C-statistic \citep{Cash79}. The small number of counts prevented us from determining the temperature of accretion spots, so it was fixed at a typical value of $25$~keV. An X-ray flux in the range  $0.1-10$~keV of $F_X = (2.4 \pm 0.3) \times 10^{-13}$~erg/cm$^{2}$/s, corrected for interstellar absorption, was obtained. Given the parallax $p'' = (4.21 \pm 0.24)\times 10^{-3}$ \citep{Gaia20}, we have an X-ray luminosity of $L_X =  (1.6 \pm 0.3)\times 10^{31}$~erg/s. This corresponds to an accretion rate onto the white dwarf of $\dot{M} \sim L_X R_1/G M_1 \approx 10^{-12}$~M$_{\odot}$/year.

\section{DISCUSSION AND CONCLUSION}

The long-term light curve of {\obj} shows two types of outbursts. One is a typical superoutburst with an amplitude of $\approx 6$~mag. This is a boundary amplitude between superoutbursts  of cataclysmic variables of the SU~UMa and WZ~Sge types. From the densest series of ZTF observations, there were most likely no superoutbursts in the interval of $\sim 5$ years. This suggests that {\obj} is a WZ~Sge type system. The two modest outbursts following the superoutburst do not resemble either normal outbursts of SU~UMa type systems (normal outbursts usually occur almost regularly) or rebrightenings in ``classical'' WZ~Sge type stars (rebrightenings are usually observed during the decline of the superoutburst, here $\approx 10$~days after the end of the superoutburst). An interesting feature of {\obj} is the $0.5$~mag higher brightness after the superoutburst compared to the pre-superoutburst luminosity. This phenomenon seems to be common among WZ~Sge type stars and has been observed, for example, in V455~And \citep{Tovmassian22}, LS~And \citep{Kato23}, KSN:BS-C11a \citep{Ridden19}, BW~Sql \citep{Neustroev23} (see also similar phenomena in SU~UMa-type stars in \cite{Pavlenko14, Pavlenko24}). The orbital period of {\obj} $P_{orb} = 0.05994478(4)$~days ($\approx 86.32$~min) agrees well with the period distribution of WZ~Sge type stars, where most systems have periods less than $0.06$~days \citep{Kato15}.

The light curves of {\obj} show a variability with a period of $P_{s}\approx 6.45$~min, which has been stable for at least the last 4 years. The stability of the period indicates that it is related to the rotation of the white dwarf rather than to non-radial pulsations, allowing {\obj} to be classified as an intermediate polar. The double-humped shape of the phase light curve indicates a two-pole accretion on the white dwarf. The ratio of the rotation period of the white dwarf to the orbital period $P_s/P_{orb} \approx 0.075$ belongs to the range $0.01-0.1$, where most of the known intermediate polars are located. The outbursts also make {\obj} a member of a rare subclass of outbursting intermediate polars (see Fig. 8 in \citealt{Pavlenko19}). 

{\obj} also exhibits out-of-eclipse variability modulated with the orbital period. The light curves convolved with the orbital period have a double-humped shape (outside the eclipse). This light behavior can be interpreted by the effects of the ellipsoidality of the donor with temperature $T\sim 2500$~K on the derived white dwarf parameters. However, given the uncertainties in the system parameters, variability of a different nature should not be excluded. For example, a similar double-humped brightness modulation is observed among WZ~Sge-type stars with low-mass donors, where the donor luminosity is insufficient to reproduce the observed brightness amplitude. Three interpretations of this phenomenon can be mentioned. The first one consists in changes in the visibility conditions of the hot spot (hot line) formed by the interaction of the accretion stream with the transparent accretion disk. In the second case, double-humped variability is associated with a spiral structure in the accretion disk that arises at the $2:1$ resonance \citep{Alives10, Zharikov13}. The $2:1$ resonance can only occur in the outer parts of the accretion disk at a mass ratio $q \le 0.1$. This is consistent within errors with the estimate of $q$ for {\obj}. Another interpretation of this phenomenon is proposed by \cite{Kononov15} and consists in the interaction of a precessing density wave with shock regions in the disk.

From the analysis of the eclipse profile the mass ratio $q = 0.11 \pm 0.02$ and the orbital inclination $i = 83.8 \pm 1.1^{\circ}$ were estimated. The eclipse egress/ingress duration corresponds to a white dwarf mass of $M_1 = 0.66 \pm 0.06$~M$_{\odot}$. The mass of the donor $M_2 = 0.073 \pm 0.015$~M$_{\odot}$ is close to the mass $M_{bounce} = 0.063^{+0.005}_{-0.002}$~M$_{\odot}$ at which the evolution of the cataclysmic variable changes from a decreasing period to an increasing period (\citealt{McAllister19}, see also \citealt{Knigge11}). However, it should be noted that the white dwarf mass estimate is model dependent and can be affected by the contribution of the accretion disk radiation, the shift of the hot spot brightness center from the ballistic trajectory, and the presence of accretion spots on the surface of the white dwarf. Since the estimate of the disk contribution from the light curve is small (not more than 10\%), the determination of the mass of the white dwarf in our work was performed neglecting the possible disk contribution.
The white dwarf temperature $T \approx 13000$~K is common for cataclysmic variables with orbital periods $P\sim 80-90$~min \citep{Townsley09, Pala22, Knigge11}. Doppler tomograms of {\obj} show the presence of a hot spot formed by the interaction between the accretion stream and the accretion disk. The tomograms also show that the accretion disk emission forms within the $3:1$ resonance radius. The orbital motion of the inner parts of the accretion disk is not distinguished from the radial velocity curves ($K\lesssim 50$~km/s), which is consistent with a low mass ratio of the system.

The low X-ray luminosity $L_X = (1.6 \pm 0.3)\times 10^{31}$~erg/s allows us to assign {\obj} to the small group of low luminosity intermediate polars\footnote{See also https://asd.gsfc.nasa.gov/Koji.Mukai/iphome/catalog/llip.html} satisfying the $L_X<10^{33}$~erg/s constraint \citep{Pretorius14}. The recently discovered intermediate polar SRGe J194401.8+28445 \citep{Kolbin24} probably belongs to the same group. The X-ray luminosity found is consistent with the luminosity of WZ~Sge type stars in the quiescent state \citep{Amantayeva21, Neustroev23, Schwope24}. A low accretion rate on the white dwarf $\dot{M} \lesssim 10^{-12}$ ~M$_{\odot}$/year is expected for WZ~Sge-type stars in the quiescent state \citep{Neustroev23}. However, it should be noted that the X-ray luminosity in the quiescent state poorly reflects the mass transfer rate in the system, since the accretion rate in the accretion disk decreases as we approach the white dwarf \citep{Cannizzo93}. 

From the results of this work, we can conclude that {\obj} is an interesting member of the dwarf novae with a magnetized white dwarf. {\obj} complements the small number of eclipsing WZ~Sge type stars, making this system an important object for understanding the evolutionary status of WZ~Sge type stars.

{\bf Acknowledgments.} The study was supported by the Russian Science Foundation grant No. 22-72-10064, https://rscf.ru/project/22-72-10064/. Observations with the SAO RAS telescopes are supported by the Ministry of Science and Higher Education of the Russian Federation. The renovation of telescope equipment is currently provided within the national project ``Science and universities''.

\end{document}